\documentstyle [12pt]{article}
\makeatletter \oddsidemargin  0in \evensidemargin 0in
\textwidth16cm \RequirePackage[dvips]{graphicx} \textheight 20.5cm
\newcommand{\singlespacing}{\let\CS=\@currsize\renewcommand{\baselinestretch}{1.5}\tiny\CS}
\makeatletter \oddsidemargin  -.1in \evensidemargin -.1in
\newcommand{\doublespacing}{\let\CS=\@currsize\renewcommand{\baselinestretch}{1.35}\tiny\CS}

\def\@citex[#1]#2{\if@filesw\immediate\write\@auxout{\string\citation{#2}}\fi
  \def\@citea{}\@cite{\@for\@citeb:=#2\do
    {\@citea\def\@citea{,\linebreak[0]\hskip0pt plus .2em}%
      \@ifundefined{b@\@citeb}%
    {{\bf ?}\@warning{Citation `\@citeb' on page \thepage\space undefined}}%
      \hbox{\csname b@\@citeb\endcsname}}}{#1}}

\newtheorem{rule-def}[theorem]{Rule}
\begin{document}
\title{ Revisiting impossible quantum operations using principles of no-signalling  and non increase of entanglement under LOCC }
\author{I.Chakrabarty$^{a}$\thanks{ indranilc@indiainfo.com },S.Adhikari$^{b}$\thanks{ satyyabrata@yahoo.com }, B.S.Choudhury $^{b}$
\\
$^{a}$Heritage Institute of Technology,Kolkata-107,West Bengal,India\\
$^{b}$ Bengal Engineering and Science
University,Howrah-711103,West Bengal,India}
\date{}
\maketitle{}
\begin{abstract}
In this letter, we show the impossibility of the general operation
introduced by Pati [3] using two different but consistent
principles (i) no-signalling (ii) non increase of entanglement
under LOCC.

\end{abstract}
The rules of quantum mechanics makes certain processes impossible.
Neither we can clone an quantum state [1] nor we can delete one of
the two identical copies of a arbitrary quantum state [2].
Operations like cloning, conjugation, complementing etc comes
under the broad heading 'General Impossible operation' [3]. In the
recent past impossibility of operations like cloning ,flipping and
deletion had been shown from two fundamental principles namely (i)
principle of no-signalling and(ii) non increase of entanglement
under LOCC [4,5,6].

In this letter we show that 'General Impossible Operations'[3]
which will act on the tensor product of an unknown quantum state
and blank state at the input port to produce the original state
along with a function of the original state at the output port is
not feasible in the quantum world from two basic principles: No
signalling principle and Thermodynamical law of Entanglement.
\\

Suppose there is a singlet state consisting of two particles
shared by two distant parties Alice and Bob. The state is given
by,
\begin{eqnarray}
|\chi\rangle_{12}=\frac{1}{\sqrt{2}}(|0\rangle|1\rangle-|1\rangle|0\rangle)\nonumber\\
=\frac{1}{\sqrt{2}}(|\psi\rangle|\overline{\psi}\rangle-|\overline{\psi}\rangle|\psi\rangle)
\end{eqnarray}
where $ \{ |\psi\rangle, |\overline{\psi}\rangle \} $ are mutually
orthogonal spin states or in other words they are mutually
orthogonal polarizations in case of photon particles. Alice is in
possession of the first particle
and Bob is in possession of the second particle.\\
No-signalling principle states that if one distant partner (say,
Alice) measures her particle in any one of the two basis namely $
\{ |0\rangle, |1\rangle \} $ and $ \{ |\psi\rangle,
|\overline{\psi}\rangle \} $, then measurement outcome of the
other party (say, Bob) will remain invariant. At this point one
might ask an interesting question: Is there any possibility for
Bob to know the basis in which Alice measures her qubit, if he
apply the operations defined as 'General Impossible operation'[1] on his qubit.\\
Let us consider a situation where Bob is in possession of a
hypothetical machine whose action in two different basis $ \{
|0\rangle, |1\rangle \} $ and $ \{ |\psi\rangle,
|\overline{\psi}\rangle \} $ is defined by the transformation,
\begin{eqnarray}
|i\rangle|\Sigma\rangle\longrightarrow|i\rangle|F(i)\rangle~~~~(i=0,1)\\
|j\rangle|\Sigma\rangle\longrightarrow|j\rangle|F(j)\rangle~~~~(j=\psi,\overline{\psi})
\end{eqnarray}
where $\{|\Sigma\rangle\}$ is the ancilla state attached by Bob .
These set of transformations
was first introduced by Pati in [1].\\
After the application of the transformation defined in (2-3) by
Bob on his particle, the singlet state takes the form
\begin{eqnarray}
|\chi\rangle|\Sigma\rangle\rightarrow\frac{1}{\sqrt{2}}(|0\rangle|1\rangle|F(1)\rangle-|1\rangle|0\rangle|F(0)\rangle)\nonumber\\
=\frac{1}{\sqrt{2}}(|\psi\rangle|\overline{\psi}\rangle|F(\overline{\psi})\rangle-|\overline{\psi}\rangle|\psi\rangle|F(\psi)\rangle)
\end{eqnarray}
Now Alice can measure her particle in two different basis. If
Alice measures her particle in $ \{|0\rangle, |1\rangle \} $, then
the reduced density matrix describing Bob's subsystem is given by,
\begin{eqnarray}
\rho_B= \frac{1}{2}[|1\rangle\langle1|\otimes|F(1)\rangle\langle
F(1)|+ |0\rangle\langle0|\otimes|F(0)\rangle\langle F(0)|~]
\end{eqnarray}
On the other hand if Alice measures her particle in the basis $ \{
|\psi\rangle, |\overline{\psi}\rangle \} $ then the state
described by the reduced density matrix in the Bob's side is given
by,
\begin{eqnarray}
\rho_B=
\frac{1}{2}[|\overline{\psi}\rangle\langle\overline{\psi}|\otimes
|F(\overline{\psi})\rangle\langle
 F(\overline{\psi})|+ |\psi\rangle\langle\psi|\otimes |F(\psi)\rangle\langle F(\psi)|~]
\end{eqnarray}
Since the statistical mixture in (5) and (6) are different, so
this would have allow Bob to distinguish the basis in which Alice
has performed the measurement and this lead to superluminal
signalling. But this is not possible from the principle of
'no-signalling', so we arrive at a contradiction. Hence, we
conclude from the principle of
no-signalling that the transformation defined in (2-3) is not possible in the quantum world.\\\\
Now we will show that the operations referred to as 'General
impossible operations' is not feasible in the quantum world from
the principle of non increase of entanglement under LOCC [ in
general one can only claim that it is conserved under a bilocal unitary operation ].\\
Let Alice and Bob share an entangled state
\begin{eqnarray}
|\Psi\rangle_{AB}=\frac{1}{\sqrt{2}}[|0\rangle|\psi_1\rangle +
|1\rangle|\psi_2\rangle]|\Sigma\rangle
\end{eqnarray}
If Bob operates 'General Impossible operations' on his qubit, then
the entangled state takes the form
\begin{eqnarray}
|\Psi\rangle^I_{AB}=\frac{1}{\sqrt{2}}[|0\rangle|\psi_1\rangle|F(\psi_1)\rangle
+ |1\rangle|\psi_2\rangle|F(\psi_2)\rangle]
\end{eqnarray}
It can be easily shown that the reduced density matrix describing
Alice's subsystem before and after the physical operation will be
different. The reduced density matrices before and after the
application of general operations are given by,
\begin{eqnarray}
\rho_A=\frac{1}{2}[|0\rangle\langle0|+|1\rangle\langle1|+|0\rangle\langle1|(\langle\psi_2|\psi_1\rangle)
+|1\rangle\langle0|(\langle\psi_1|\psi_2\rangle)]
\end{eqnarray}
and
\begin{eqnarray}
\rho^I_A&=&\frac{1}{2}[|0\rangle\langle0|+|1\rangle\langle1|+|0\rangle\langle1|(\langle\psi_2|\psi_1\rangle)(\langle
F(\psi_2)|F(\psi_1)\rangle)
+{}\nonumber\\&&|1\rangle\langle0|(\langle\psi_1|\psi_2\rangle)(\langle
F( \psi_1)|F(\psi_2)\rangle)]
\end{eqnarray}
The largest eigenvalues corresponding to these density matrices
are given by, $\lambda=\frac{1}{2}+\frac{|\alpha|}{2}$ and
$\lambda^I=\frac{1}{2}+\frac{\sqrt{\alpha^2\overline{\alpha}^2}}{2}$
respectively, where $\alpha =\langle\psi_1|\psi_2\rangle=\langle
F(\psi_1)|F(\psi_2)\rangle$ and
$|\psi_{1}\rangle=a|0\rangle+b|1\rangle$, $|\psi_{2}\rangle=c|0\rangle+d~ exp~(i\theta)|1\rangle$ .\\

Let us consider $\alpha=X+iY$ where $X=ac+bdcos\theta$ and $Y=bdsin\theta$.\\
From the principle of non increase of entanglement we can write,
$E(|\psi\rangle_{AB}^I)\leq E(|\psi\rangle_{AB})$, which implies $\lambda\leq\lambda^I$.
 That is, we are assuming that there is no increase of entanglement.\\
Therefore $\lambda\leq\lambda^I\Rightarrow|\alpha|^{2}\geq
1\Rightarrow X^2+Y^2\geq1\Rightarrow
cos\theta\geq\frac{1-(a^2c^2+b^2d^2)}{2abcd}$.\\
Now we observe that $(a-c)^2+(b-d)^{2}\geq0\Rightarrow ac+bd\leq
1$. Again squaring both sides, the inequality reduces to
$a^{2}c^{2}+b^{2}d^{2}+2abcd\leq1\Rightarrow\frac{1-(a^2c^2+b^2d^2)}{2abcd}\geq1\Rightarrow\cos\theta\geq1$.
The equality holds when $a=c$ and $b=d$. So we arrive at a
contradiction when $a\neq c$ and $b\neq d$. Hence
$\lambda>\lambda^I$. This violates the principle of no increase of
entanglement under LOCC. This proves the
impossibility of 'General Impossible Operations'  from the principle of non increase of entanglement under LOCC.\\\\

{\bf Acknowledgement:}\\
S.A acknowledges CSIR (project no.F.No.8/3(38)/2003-EMR-1, New
Delhi) for providing fellowship to carry out this work. I.C
acknowledges Prof C.G.Chakrabarti, N.Ganguly for their cooperation
in completing this work. The part of the work was done at the
Institute of Physics (Bhubaneswar) July-August 2005. I.C and S.A
are grateful for the institute's hospitality. \\

{\bf Reference:}\\
$[1]$ W.K.Wootters and W.H.Zurek, Nature \textbf{299},802(1982).\\
$[2]$ A.K.Pati and S.L.Braunstein, Nature \textbf{404},164(2000)\\
$[3]$ A.K.Pati, Phys.Rev.A \textbf{66},062319 (2002)\\
$[4]$ N.Gisin ,Phys .Lett.A \textbf{242},1-3 (1998)\\
$[5]$ A.K.Pati and S.L.Braunstein,Phys.Lett.A \textbf{315},208-212
(2003)\\
$[6]$ I.Chattopadhyay.et.al.Phys. Lett. A, \textbf{351}, 384-387
(2006)
\end{document}